\documentclass[%
 aip,
 amsmath,amssymb,
 reprint,%
]{revtex4-1}

\usepackage{graphicx}
\usepackage{dcolumn}
\usepackage{bm}
\usepackage{siunitx}
\usepackage{multirow}
\usepackage{caption}

\usepackage[utf8]{inputenc}
\usepackage[T1]{fontenc}
\usepackage{mathptmx}
\usepackage{etoolbox}

\makeatletter
\def\@email#1#2{%
 \endgroup
 \patchcmd{\titleblock@produce}
  {\frontmatter@RRAPformat}
  {\frontmatter@RRAPformat{\produce@RRAP{*#1\href{mailto:#2}{#2}}}\frontmatter@RRAPformat}
  {}{}
}%
\makeatother
\begin{document}

\preprint{AIP/123-QED}

\title[]{Consistent temporal behaviors over a non-Newtonian/Newtonian two-phase flow}

\author{H. H. Song}
  \altaffiliation[]{These authors contributed equally.}
\author{J. S. Zhang}
  \altaffiliation[]{These authors contributed equally.}
\affiliation{
School of Mechatronic Engineering and Automation, Shanghai University, Shanghai 200072, China
}

\author{Z. L. Wang}
  \altaffiliation[Corresponding author: ]{wng\_zh@i.shu.edu.cn}
\affiliation{
Shanghai Key Laboratory of Mechanics in Energy Engineering, Shanghai Institute of Applied Mathematics and Mechanics, School of Mechanics and Engineering Science, Shanghai University, Shanghai 200444, China
}

\date{\today}

\begin{abstract}
The temporal behaviors of non-Newtonian/Newtonian two-phase flows are closely tied to flow pattern formation mechanisms, influenced significantly by the non-Newtonian index, and exhibiting nonlinear rheological characteristics. The rheological model parameters are model-dependent, resulting in poor predictability of their spatio-temporal characteristics. This limitation hinders the development of a unified analytical approach and consistent results, confining researches to case-by-case studies. In this study, sets of digital microfluidics experiments, along with continuous-discrete phase-interchanging schemes, yielded 72 datasets under various non-Newtonian fluid solution configurations, micro-channel structures, and multiple Carreau-related models. Among these datasets, we identified consistent temporal behaviors featuring non-Newtonian flow patterns in  digital microfluidics. In the context of significant model dependency and widespread uncertainty in model parameters for non-Newtonian fluid characterization, this study demonstrates that the consistent representation of the non-Newtonian  behavior exit, and may be essentially independent of specific data or models. Which finding holds positive implications for understanding the flow behavior and morphology of non-Newtonian fluids.
\end{abstract}

\maketitle

\begin{figure*}[!htbp]
    \includegraphics{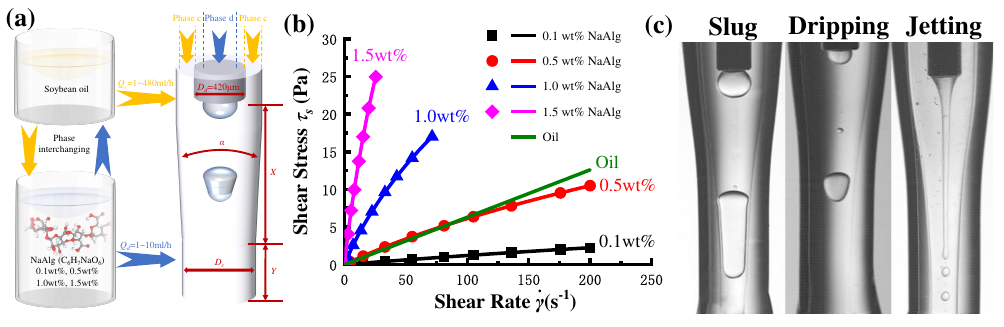}
    \captionsetup{justification=raggedright,singlelinecheck=false}
    \caption{\label{fig:setup}(a) Experimental setup and microchannel structure, highlighting phase interchanging design. (b) Rheological curves of experimental fluids, including soybean oil and NaAlg solutions in four mass fractions (\SI{0.1}{wt\%}, \SI{0.5}{wt\%}, \SI{1.0}{wt\%}, \SI{1.5}{wt\%}). (c) Flow pattern images, including slug, dripping, and jetting flow.}
\end{figure*}

In digital material fabrication and complex encapsulation material studies, non-Newtonian fluids are increasingly utilized, particularly in chemical reactions and structure formation\cite{lin2022gas, lee2023rat, wan2022phy}. Their complex rheological properties and impacts on material fabrication processes are now research focal points\cite{leu2021inv}. Understanding the role of these fluids in material formation is crucial for developing new fabrication methods and improving existing technologies\cite{kor2022phy}. Non-Newtonian fluids exhibit complex rheological properties, characterized by a nonlinear stress-strain relationship, which results in flow behavior distinct from that of Newtonian fluids. Non-Newtonian fluids also exhibit complex physical properties, such as the Weissenberg effect\cite{wei1947a} and Tom’s phenomenon\cite{vir1967the}, adding complexities and difficulties to related research. To accurately describe the behavior of non-Newtonian fluids, complex mathematical models are required. Choosing an appropriate model and determining its parameters pose a challenge. Therefore, rheological parameters like viscosity ratio, density ratio, consistency coefficient, and non-Newtonian index\cite{bar2020ste, amo2020non} have been introduced by scholars to better understand their flow behaviors. Characterized by shear-thinning/shear-thickening properties, these fluids present challenges in viscosity measurement\cite{le2023sol}. Consequently, various models such as the Power-Law model\cite{wae1923vis}, Carreau model\cite{kok1984pre}, and Herschel-Bulkley model\cite{her1926kon} are frequently used for characterizing their rheological properties. However, the critical non-Newtonian index is usually not independently included in related characterizations. Furthermore, the non-linearity and complexity of non-Newtonian fluid rheological properties hinder the establishment of unified flow behavior patterns and characterizations in experiments.

Among the typical coaxial configuration of digital microfluidic basic structures\cite{liu2016two, wan2023rif, ono2013met}, in these article, we designed multiple sets of soybean oil/sodium alginate (NaAlg) solution two-phase flow in converging microchannels\cite{zha2020exp, wan2015spe, wan2022uni} (as depicted in Fig.~\ref{fig:setup}(a)), including three microchannels, four solution configurations, and two phase interchanging experiments. While the convergence angle $\alpha $, converging section length $X$, and straight pipe length $Y$ all significantly impact the results of fluid units, the three sizes of pipe specifications we have used here are not intentionally set to possess strict regular characteristics; instead, we prefer them to be more arbitrary (as shown in Table~\ref{tab:size}). The phase interchanging designs are: A) oil as the continuous phase (Phase $c$) and NaAlg solution as the dispersed phase (Phase $d$); B) NaAlg solution as the continuous phase and oil as the dispersed phase. The NaAlg solutions were prepared in four mass fractions: \SI{0.1}{wt\%}, \SI{0.5}{wt\%}, \SI{1.0}{wt\%}, \SI{1.5}{wt\%}. As shown in Fig.~\ref{fig:setup}(b), the rheological curves of these solutions and oil revealed that NaAlg solution is shear-thinning fluid; the Carreau model, modified Carreau model (polynomial model), and Carreau-Yasuda model\cite{yas1979mas} were employed to fit the shear viscosity curves of NaAlg solutions (shown in Table~\ref{tab:model}). Flow patterns of slug, dripping, and jetting, forming monodisperse microdroplets, were observed in the experiments, shown in Fig.~\ref{fig:setup}(c). 

The varied selections of pipelines, non-Newtonian fluid solution properties, two-phase exchange configurations, and viscosity models in the preceding experiments has resulted in a large and complex data, leading to $2\times 4\times 3\times 3=72$ sets of different temporal and spatial flow pattern dataset. To simplify the issue, it would be convenient to describe these datasets as a whole. Here, we use an array element $M[i][j][k]$ to represent a specific dataset, facilitating the description and manipulation of these datasets. Where $M=A$ or $B$ represents phase interchanging scheme of A or B; $i=1\sim 3$ represents Channel-1, Channel-2, and Channel-3; $j=1\sim 3$ represents the Carreau, modified Carreau, and Carreau-Yasuda models for rheological properties; $k=1\sim 4$ represents NaAlg solution mass fractions \SI{0.1}{wt\%}, \SI{0.5}{wt\%}, \SI{1.0}{wt\%}, \SI{1.5}{wt\%}; and the symbol `$:$' represents all experimental data groups for each category.

\begin{table}[!htbp]
\captionsetup{justification=raggedright,singlelinecheck=false}
\caption{\label{tab:size} Three microchannel configurations.}
\begin{ruledtabular}
\begin{tabular}{ccccc}
Channel & $\alpha $(\SI{}{\degree}) & $D_c $(\SI{}{\mu m})  & $X$(\SI{}{\mu m})     & $Y$(\SI{}{\mu m}) \\  \hline
1       & 9                         & 1000                  & 2000                  & 4000              \\
2       & 8                         & 750                   & 1500                  & 1500              \\
3       & 8                         & 1000                  & 1000                  & 5500              \\
\end{tabular}
\end{ruledtabular}
\end{table}

The parameters in non-Newtonian fluid models are typically determined through experiments, which may introduce degrees of uncertainty. The accurate determination of parameters is crucial for the precision of simulations. However, our experiments and practical experience indicate that obtaining similar rheological parameters for non-Newtonian fluids using different models is challenging. Here, the equations for three Carreau related models are shown in Table~\ref{tab:model}, where $n$ is the non-Newtonian index, $\eta $ is apparent viscosity, $\eta_0$ is zero-shear-rate viscosity, $\eta_\infty $ is infinity-shear-rate viscosity, $\dot{\gamma } $ is shear rate, $\lambda $ is material relaxation time, and $a$ is Yasuda index. The polynomial form ($\lambda_1$, $\lambda_2$, $\lambda_3$, $\lambda_4$) is utilized in the modified Carreau model to improve fitting precision. The corresponding material properties and model fitting parameters are shown in Table~\ref{tab:parameter}, where $\rho $ is density and $\sigma $ is interfacial tension. Parameters for the Carreau model were calculated by Zhang et al.\cite{zha2022acc}. The non-Newtonian index $n$ is generally positive, few instances of $n$ is negative\cite{sur2018rhe}. For lower mass fractions (\SI{0.1}{wt\%}, \SI{0.5}{wt\%}), $n$ in the Carreau model is negative, whereas it is positive in the modified Carreau and Carreau-Yasuda models; this results reversed for higher mass fractions (\SI{1.0}{wt\%}, \SI{1.5}{wt\%}).

\begin{table*}[!htbp]
\captionsetup{justification=raggedright,singlelinecheck=false}
\caption{\label{tab:model}Equations for Carreau related models.}
\begin{ruledtabular}
\begin{tabular}{cc}
Model               & Formula                                                                                                                                                           \\  \hline
Carreau             & $\eta=(1+(\lambda\dot{\gamma})^2)^{\frac{n-1}{2}}(\eta_0-\eta_\infty)+\eta_\infty $                                                                               \\
Modified Carreau    & $\eta=(1+\lambda_1\dot{\gamma}+(\lambda_2\dot{\gamma})^2+(\lambda_3\dot{\gamma})^3+(\lambda_4\dot{\gamma})^4)^{\frac{n-1}{2}}(\eta_0-\eta_\infty)+\eta_\infty $   \\
Carreau-Yasuda      & $\eta=(1+(\lambda\dot{\gamma})^a)^{\frac{n-1}{a}}(\eta_0-\eta_\infty)+\eta_\infty $                                                                               \\
\end{tabular}
\end{ruledtabular}
\end{table*}

\begin{table*}[!htbp]
\captionsetup{justification=raggedright,singlelinecheck=false}
\caption{\label{tab:parameter}Material properties and model fitting parameters. Boldface denotes the fitting results of the non-Newtonian index $n$, which is rarely negative, playing a significant role in flow pattern classification.}
\begin{ruledtabular}
\begin{tabular}{ccccccccccccccc}
\multicolumn{5}{c}{\multirow{2}{*}{NaAlg solution properties}} & \multicolumn{10}{c}{Model parameters}                                                                               \\ \cline{6-15} 
\multicolumn{5}{c}{}                                           & \multicolumn{2}{c}{Carreau}            & \multicolumn{5}{c}{Modified Carreau}  & \multicolumn{3}{c}{Carreau-Yasuda} \\ \hline
Concentration   & $\rho $(\SI{}{kg/m^3})    & $\sigma $(\SI{}{N/m}) & $\eta_0$(\SI{}{Pa \cdot s})   & $\eta_\infty $(\SI{}{Pa \cdot s}) & $\lambda $(\SI{}{s})  & \textbf{\textit{n}}   & $\lambda_1$(\SI{}{s}) & $\lambda_2$(\SI{}{s}) & $\lambda_3$(\SI{}{s}) & $\lambda_4$(\SI{}{s}) & \textbf{\textit{n}}          & $\lambda$(\SI{}{s})   & $a$       & \textbf{\textit{n}}   \\
\SI{0.1}{wt\%}  & 989                       & 0.0223                & 0.010                         & 0.004                             & 0.272                 & \textbf{-0.583}       & 0.008                 & 0.003                 & 0.003                 & 0                     & \textbf{0.492}    & 0.002                             & 0.847     & \textbf{0.607}        \\
\SI{0.5}{wt\%}  & 999                       & 0.0231                & 0.057                         & 0.005                             & 0.284                 & \textbf{-0.567}       & 0.010                 & 0.017                 & 0                     & 0.005                 & \textbf{0.719}    & 0.008                             & 1.183     & \textbf{0.612}      \\
\SI{1.0}{wt\%}  & 1007                      & 0.0245                & 0.380                         & 0.062                             & 0.467                 & \textbf{0.162}        & 0.017                 & 0                     & 0                     & 0.003                 & \textbf{-0.585}   & 0.003                             & 0.655     & \textbf{-0.232}     \\
\SI{1.5}{wt\%}  & 1017                      & 0.0250                & 1.181                         & 0.142                             & 0.165                 & \textbf{0.437}        & 0.041                 & 0                     & 0                     & 0.004                 & \textbf{-0.741}   & 0.002                             & 0.537     & \textbf{-0.611} 
\end{tabular}
\end{ruledtabular}
\end{table*}

\begin{figure*}[!htbp]
    \includegraphics{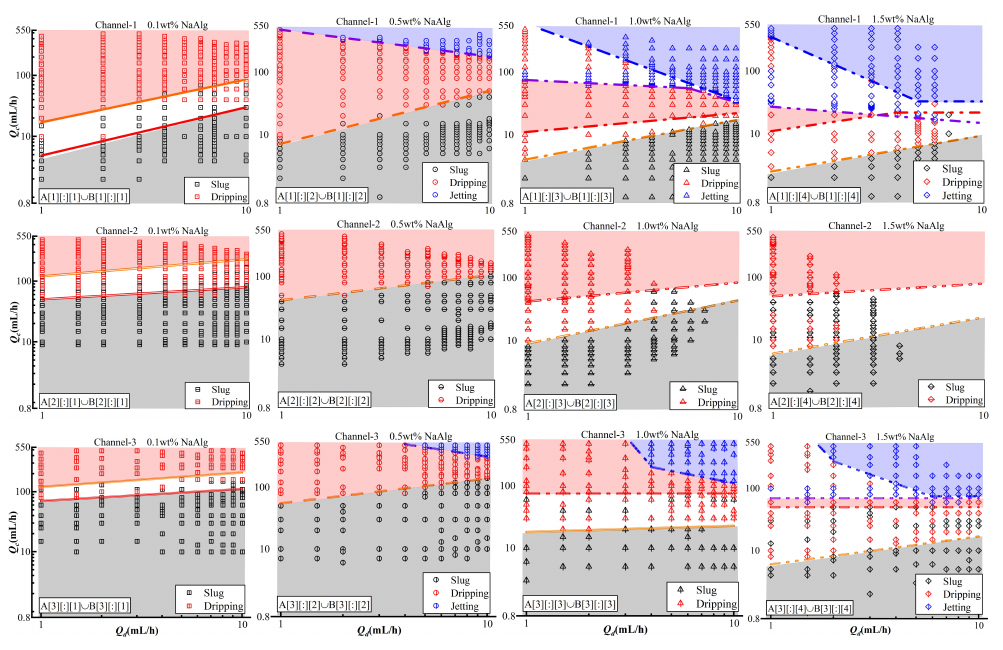}
    \captionsetup{justification=raggedright,singlelinecheck=false}
    \caption{\label{fig:flow_map}$Q_c\sim Q_d$ flow pattern map, including three microchannel structures, two experimental designs, and four mass fraction conditions. Flow patterns of $A[i][:][k]$ and $B[i][:][k]$ are overlaid for illustration. The slug-dripping transition lines in $A[i][:][k]$ and $B[i][:][k]$ are marked by red and orange lines, respectively, while blue and purple lines indicate the dripping-jetting transition lines in $A[i][:][k]$ and $B[i][:][k]$. Shaded regions signify identical flow patterns in $A[i][:][k]$ and $B[i][:][k]$, with slug, dripping, and jetting patterns represented by black, red, and blue shades respectively. Coinciding transition lines are observed for $A[i][:][2]$ and $B[i][:][2]$ flow patterns. Unshaded areas highlight different flow patterns between $A[i][:][k]$ and $B[i][:][k]$.}
\end{figure*}

\begin{figure*}[!htbp]
    \includegraphics{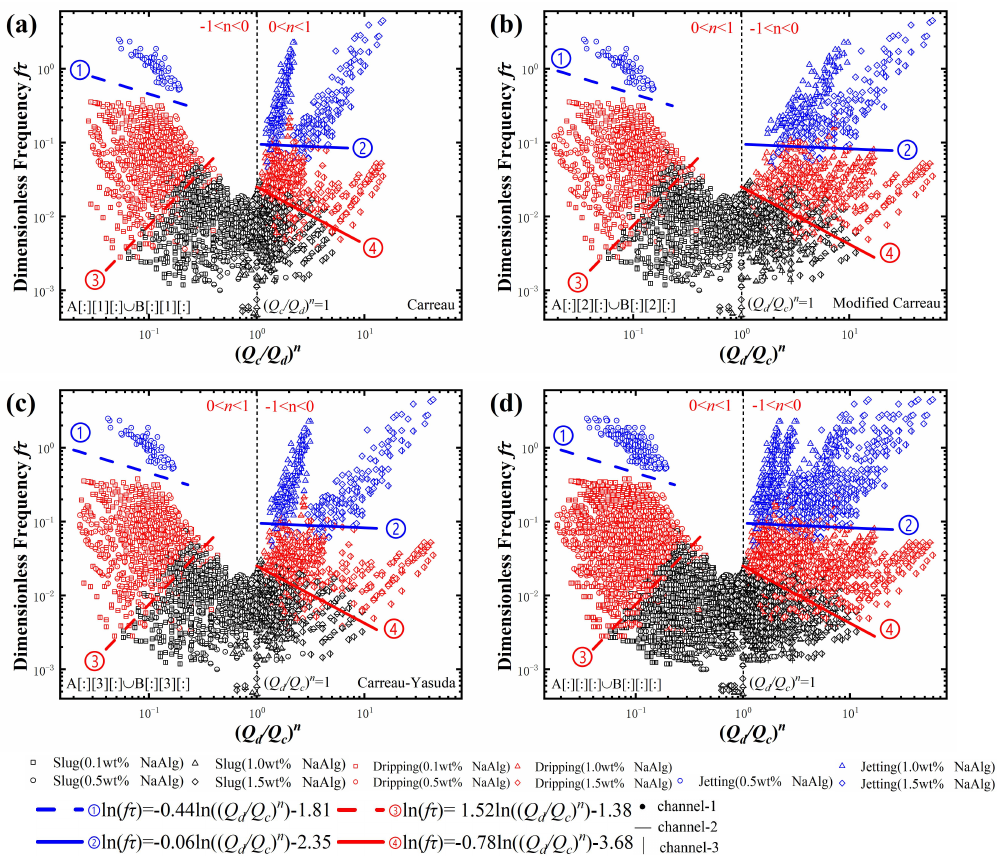}
    \captionsetup{justification=raggedright,singlelinecheck=false}
    \caption{\label{fig:butterfly}The flow rate spatial distribution map of temporal characteristic, incorporating the effect of the non-Newtonian index $n$. Transition lines are calculated by machine learning algorithm using MATLAB (fitcsvm; Support Vector Machines). The red line represents the slug-dripping transition, and the blue line indicates the dripping-jetting transition. To differentiate flow pattern results from three microchannel structures, symbols of different concentrations and flow patterns are augmented with dots, horizontal, and vertical lines. (a) $f \tau \sim\left(Q_c / Q_d\right)^n$ phase diagram for $A[:][1][:]$ and $B[:][1][:]$ flow patterns. (b) $f \tau \sim\left(Q_d / Q_c\right)^n$ phase diagram for $A[:][2][:]$ and $B[:][2][:]$ flow patterns, and (c) $A[:][3][:]$ and $B[:][3][:]$ flow patterns. (d) $f \tau \sim\left(Q_d / Q_c\right)^n$ phase diagram for all $A[:][:][:]$ and $B[:][:][:]$ flow patterns, where the results from $A[:][1][:]$ and $B[:][1][:]$ flow patterns are after reflected about the symmetry axis $\left(Q_d / Q_c\right)^n=1$.}
\end{figure*}

Flow patterns can be characterized by velocity\cite{fu2015flow} or flow rate\cite{wan2022uni}, facilitating clear flow pattern transition lines. If plotting $Q_c\sim Q_d$ flow pattern maps using corresponding continuous-dispersed phase interchanging data $A[i][:][k]\cup B[i][:][k]$ (where $Q_d$ and $Q_c$ represent dispersed and continuous phase flow rate respectively), as a single graph, as shown in Fig.~\ref{fig:flow_map}. The overlapping sections of the $A[i][:][2]$ and $B[i][:][2]$ flow pattern distributions within the shaded areas are identified, and distinct flow patterns are differentiated using colors; if $A[i][:][k](k\neq 2)$ and $B[i][:][k](k\neq 2)$ flow pattern distributions differ, the areas are left blank. Observations reveal that despite the complex variability in flow pattern transition boundaries across all results, overall, the flow pattern maps exhibit a continuous change in response to increasing concentrations of the non-Newtonian NaAlg solution. Within the assessed range of flow rate parameters, there is an expansion in the jetting region, contrasted by a contraction in the slug and dripping regions\cite{kal2022eff}; however jetting is absent in the narrower channel-2, consistent with the more stable flow characteristic of narrower microchannels\cite{raj2019a}. Similarly, we also find that as the concentration of the non-Newtonian NaAlg solution increases, the differences in the flow pattern maps using corresponding continuous-dispersed phase interchanging data $A[i][:][k]$ and $B[i][:][k]$, shown as the blank spaces in the maps, demonstrate a decreasing and then increasing trend. This implies that there are scenarios where the flow pattern maps remain unchanged after two phase interchanging, controllable by the non-Newtonian NaAlg solution concentration, exemplified by the \SI{0.5}{wt\%} data shown as the second column of Fig.~\ref{fig:flow_map}. However, clearly, due to the differences in the flow pattern maps resulting from the phase interchanging, which are both objective and physical, as well as nonlinear and complex, finding unified transition boundaries for these data is difficult or perhaps impossible. Consequently, achieving results that surpass case-by-case analysis becomes exceedingly difficult or impossible with variations in experimental conditions or control variables.

However, in the pursuit of structural information across extensive data, it was unexpectedly found that achieving a unified spatial distribution classification is inherently difficult, whereas such feasibility is present in the temporal dimension.

The frequency of microdroplet generation is a key aspect of microdroplet dynamics and flow pattern analysis\cite{liu2018for, hua2017ac, kum2022ins}. The instances of the first $t_1$ and $i$-th $t_i$ droplets passing through a fixed position in the microchannel are recorded to calculate the droplet frequency $f=\frac{i-1}{t_i-t_1}$; the capillary time $\tau=\sqrt{\rho D_c^3/\sigma}$ refers to the time required for surface tension to influence dispersed phase droplets\cite{du2018bre}, facilitated the dimensionless droplet frequency $f\tau $. We found that to consider the characteristics of non-Newtonian fluids, it is necessary to introduce characteristic parameters of non-Newtonian fluids, among which the non-Newtonian fluid index $n$ is one of the most important parameters. By introducing non-Newtonian fluid index $n$ and plotting the flow pattern phase diagrams of all $A[:][j][:]\cup B[:][j][:]$ datasets in $f \tau{\sim}(Q_d/Q_c)^n$ space, we achieved a uniform distribution of these data in time domain, as shown in Fig.~\ref{fig:butterfly}. Flow patterns, categorized by $n$, are symmetrically divided by $(Q_d/Q_c)^n=1$ ($Q_d=Q_c$ or/and $n=0$). Significantly, the Carreau model in $f \tau{\sim}(Q_c/Q_d)^n$ phase diagram shows a flow pattern distribution that precisely matches those of other models in $f \tau{\sim}(Q_d/Q_c)^n$ phase diagrams.

Fig.~\ref{fig:butterfly} comprises four subfigures, with Fig.~\ref{fig:butterfly}(a)-(c) representing 24 sets of phase interchanging data corresponding to the Carreau model, modified Carreau model (polynomial model), and Carreau-Yasuda model, respectively; Fig.~\ref{fig:butterfly}(d) overlays the results of the previous three figures, presenting a comprehensive view of 72 data sets. Generally, the transition among slug, dripping, and jetting modes is primarily influenced by competition between the shear stress exerted on the emerging droplet by the continuous phase and the interfacial stresses\cite{nie2008emu}; increased shear stress is inclined to generate smaller droplets; according to the principle of mass conservation, smaller droplets indicate higher frequencies, which should be positioned higher in our phase diagram space, thus our results are aligning with physical expectations\cite{bai2021gen}. Remarkably, these data exhibit significant consistency in temporal behavior. The experimental results reveal a surprising uniformity across diverse conditions, including variations in microchannel types, non-Newtonian fluid viscosities from different solution configurations, and continuous-dispersed phase interchanging. Even when applying different stress-strain models, the non-Newtonian index $n$ in Table~\ref{tab:parameter} displays both positive and negative values, reflecting significant variations in the introduced non-Newtonian index; despite these variations and even reversal in sign when employing different models, the stability of the temporal behavior distribution remains unaffected, which is intriguing. This stability, unaffected by experimental variables such as microchannel structures, solution configurations, continuous-dispersed phase interchanging, and various non-Newtonian fluid constitutive relations, thus provides a unified description and a potential tool for comprehending the flow behavior and morphology of non-Newtonian fluids.

In short, this study presents a unified method and result for characterizing the temporal behaviors of soybean oil/NaAlg solution two-phase flows in converging coaxial microchannels. The robust adaptability of the method and temporal behaviors are validated through 72 diverse data sets encompassing various microchannel designs, rheological models, solution concentrations, and phase interchanging experiments. Considering the inherent nonlinear behavior of non-Newtonian fluids, attaining consistent fluid behavior is a significant challenge. The study's outcomes and characterizations provide critical insights and guidance for non-Newtonian fluid behavior research, with practical significance for material fabrication and chemical reaction control.

\section*{Acknowledgements}
The author acknowledges the National Natural Science Foundation of China (No. 11832017 and 11772183). 

The data that support the findings of this study are available from the corresponding author upon reasonable request. The authors have no conflicts to disclose.

\section*{Reference}
\nocite{*}
\bibliography{ref}

\end{document}